\documentclass[epjCONF, onecolumn,usenatbib]{svjour}
\usepackage{graphics}
\usepackage[varg]{txfonts} 
\usepackage[latin1]{inputenc}

\newcommand{\zr}{{\rm z_{reion}}}

\newcommand{\nicefrac}[2]{\leavevmode\kern.1em
            \raise.5ex\hbox{\the\scriptfont0 #1}\kern-.1em
      /\kern-.15em\lower.25ex\hbox{\the\scriptfont0 #2}}

\session-title{Conference Title, to be filled}
\begin{document}
\title{Constraining local UV field geometry at reionization using Milky Way satellites.}
\author{P. Ocvirk\inst{1}\fnmsep\thanks{\email{pierre.ocvirk@astro.unistra.fr}} \and D. Aubert\inst{1}}
\institute{Observatoire Astronomique de Strasbourg, 11 rue de l'universit\'e, Strasbourg, France}
\abstract{We present a new semi-analytical model of the population of satellite galaxies of the Milky Way, aimed at estimating the effect of the geometry of reionization at galaxy scale on the properties of the satellites. In this model reionization can be either: (A) externally-driven and uniform, or (B) internally-driven, by the most massive progenitor of the Milky Way. In the latter scenario the propagation of the ionisation front and photon dilution introduce a delay in the photo-evaporation of the outer satellites' gas with respect to the inner satellites. As a consequence, outer satellites experience a longer period of star formation than those in the inner halo. We use simple models to account for star formation, the propagation of the ionisation front, photo-evaporation and observational biases. Both scenarios yield a model satellite population that matches the observed luminosity function and mass-to-light ratios. However, the predicted population for scenario (B) is significantly more extended spatially than for scenario (A), by about 0.3 dex in distance, resulting in a much better match to the observations. 
The survival of the signature left by the local UV field during reionization on the radial distribution of satellites makes it a promising tool for studying the reionization epoch at galaxy scale in the Milky Way and nearby galaxies resolved in stars with forthcoming large surveys.} 
\maketitle
\section{Introduction}
\label{intro}
Several recent semi-analytical models (SAMs) have shown that the observed luminosity function of the satellites of the Milky Way (hereinafter MW) can be well reproduced under a number of simple assumptions for the baryonic physics. Most past studies implemented a uniform, instantaneous reionization by simply suppressing star formation in the sub-haloes below $\zr$. 
Here, on the contrary, we investigate the effect of varying the structure of the UV field at reionization on the properties of the satellites of the MW. 

\section{Model}
\label{sec:model}
Our SAM is partly inspired by \cite{busha2010}, but it describes the reionization epoch in more detail: instead of assuming instantaneous quenching of star formation, we compute photo-evaporation times using the formula of \cite{iliev2005}, as a function of halo mass and incident UV flux. Star formation shuts down in a halo once its photo-evaporation is achieved. Our model (A) represents an external reionization scenario, where all halos see the same flux. Our model (B) represents an inside-out reionization scenario, driven by an internal source. We approximate this regime by putting one central UV source at the position of the most massive progenitor of the MW. Finally, we compute the distance and luminosity-dependent detection probability of our model satellite population following the prescriptions of \cite{koposov2008}. See  \cite{ocvirk2011} for more details.

\section{Results}
\label{sec:results}
\begin{figure}
\begin{center}
\resizebox{0.6\columnwidth}{!}{\includegraphics{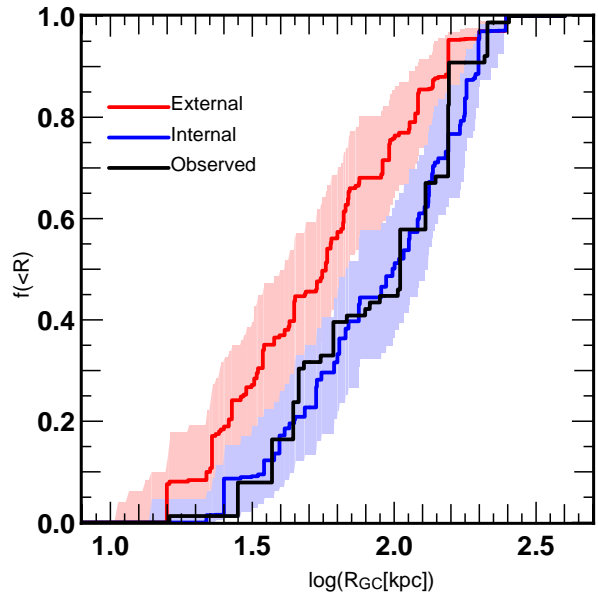}}
\end{center}
\caption{Satellites cumulative radial distribution profiles, observed (black), and from the internal and external reionization scenarios. The shaded areas show the dispersion of profiles obtained by 1000 realisations of a mock DR5 survey.}
\label{f:CRD}
\end{figure}

Both models give a very similar satellites luminosity function, and agree well with the observed one.
However fig. \ref{f:CRD} shows that there is a strong difference between the predicted cumulative radial distribution of satellites for the 2 scenarios:  the distribution for the internal reionization model is shifted outwards by about 0.3 dex with respect to the external reionization model. This is the signature of internal reionization: in the galacto-centric reionization scenario the satellites of the inner halo see a more intense UV flux than their outer halo cousins, and therefore evaporate faster. Thus the outer halo satellites experience a longer star formation activity period and  end up brighter at $z=0$. 

\section{Conclusions}
The sensitivity of the radial distribution of satellites to the structure of the UV field during reionization makes it an interesting probe for the study of reionization at galaxy scale with Pan-STARRS and upcoming LSST. Provided that the VLII halo is representative of the MW, our result suggests that star formation at the center of our Galaxy is responsible for the photo-evaporation of its own satellites and the intervening gas.


\end{document}